\documentstyle[12pt]{article}
\newcommand{\be}{\begin{equation}}
\newcommand{\ee}{\end{equation}}

\newcommand{\bear}{\begin{eqnarray}}
\newcommand{\eear}{\end{eqnarray}}
\newcommand{\mod}{\left|\hspace{-0.4cm}\begin{array}{c}\\
\end{array}\right.} \def\Hu{H_1^{0r}}
\newcommand{\parl}{\left(\hspace{-0.4cm}\begin{array}{c}\\
\end{array}\right.} \def\Hu{H_1^{0r}}
\newcommand{\parr}{\left)\hspace{-0.4cm}\begin{array}{c}\\
\end{array}\right.} \def\Hu{H_1^{0r}}
\newcommand{\corl}{\left[\hspace{-0.4cm}\begin{array}{c}\\
\end{array}\right.} \def\Hu{H_1^{0r}}
\newcommand{\corr}{\left]\hspace{-0.4cm}\begin{array}{c}\\
\end{array}\right.} \def\Hu{H_1^{0r}}
\newcommand{\llavl}{\left\{\hspace{-0.4cm}\begin{array}{c}\\
\\ \end{array}\right.} \def\Hu{H_1^{0r}}
 \def\Hu{H_1^{0r}}
\def\Hd{H_2^{0r}}
\def\hu{h_1^{0r}}
\def\hd{h_2^{0r}}
\def\mtu{m_{\tilde{t_1}}^2}
\def\mtd{m_{\tilde{t_2}}^2}
\def\mbu{m_{\tilde{b_1}}^2}
\def\mbd{m_{\tilde{b_2}}^2}
\def\IJMPA #1 #2 #3 {Int.~J.~Mod.~Phys.~{\bf A#1}\ (19#2) #3}
\def\MPLA #1 #2 #3 {Mod.~Phys.~Lett.~{\bf A#1}\ (19#2) #3}
\def\NPB #1 #2 #3 {Nucl.~Phys.~{\bf B#1}\ (19#2) #3}
\def\PLB #1 #2 #3 {Phys.~Lett.~{\bf B#1}\ (19#2) #3}
\def\PR #1 #2 #3 {Phys.~Rep.~{\bf#1}\ (19#2) #3}
\def\PRD #1 #2 #3 {Phys.~Rev.~{\bf D#1}\ (19#2) #3}
\def\PTP #1 #2 #3 {Prog.~Theor.~Phys.~{\bf #1}\ (19#2) #3}
\def\PRL #1 #2 #3 {Phys.~Rev.~Lett.~{\bf#1}\ (19#2) #3}
\def\RMP #1 #2 #3 {Rev.~Mod.~Phys.~{\bf#1}\ (19#2) #3}
\def\ZPC #1 #2 #3 {Z.~Phys.~{\bf C#1}\ (19#2) #3}

\begin{document}

\begin{titlepage}

\title{\bf The Supersymmetric Singlet Majoron Model and
 the General Upper Bound on the Lightest Higgs Boson Mass}

\author{
{\bf J.R. Espinosa} \thanks{Supported by the Alexander-von-Humboldt
Stiftung.} \\ Deutsches Elektronen Synchrotron DESY. \\
Notkestrasse 85.\ \ 22603 Hamburg. Germany}

\date{}
\maketitle
\vspace{1.5cm}
\def\baselinestretch{1.15}
\begin{abstract}
An upper bound on the tree-level mass of the lightest Higgs boson
of the Supersymmetric Singlet Majoron Model is obtained. Contrary
to some recent claims, it is shown to be of the same form as the
general mass bound previously calculated for supersymmetric models
with an extended Higgs sector. Soft-breaking masses or exotic vacuum
expectation values do not enter in the tree-level bound [which is
only controlled by the electroweak scale ($M_Z$)] and also decouple
from the most important radiative corrections to the bound (the ones
coming from the top-stop sector). The derivation of the upper bound
for general Supersymmetric Models is reviewed in order to clarify
its range of applicability.
\end{abstract}
\vspace{2cm}
\leftline{March 1995}

\thispagestyle{empty}

\vskip-19.cm
\rightline{{\bf DESY 95-037}}
\rightline{{\bf IEM--FT--102/95}}
\vskip3in

\end{titlepage}

\def\baselinestretch{1.1}

The search for a Higgs boson is one of the most challenging goals for
existing and planned accelerators. The discovery of such a fundamental
scalar would be the first step in understanding the elusive
mechanism of electroweak symmetry breaking. In the framework of
supersymmetric theories the Higgs sector is particularly constrained
and provides a unique ground for checking whole classes of supersymmetric
models. In contrast with the arbitrariness in the masses of most of the new
particles predicted by supersymmetry (which are only weakly restricted by
naturalness criteria) it seems to be a general feature of
Supersymmetric Standard Models the presence of a light Higgs particle in
the spectrum (with mass of order $M_Z$ even in the limit of unnaturally large
supersymmetric masses).

As is well known, in the Minimal Supersymmetric Standard Model the tree
level mass $m_h$ of the lightest Higgs boson is bounded by
$M_Z | \cos2\beta |$.
Radiative corrections to the mass of this Higgs boson can be large if
the mass of top and stops is large, and the tree level bound can be spoiled
\cite{radcor}. After including next-to-leading log corrections \cite{radsub}
the numerical bound $m_h<140\ GeV$ (for a top mass below $190\ GeV$ and
stops not heavier than $1\ TeV$) is found, so that, even if the lightest
Higgs can scape detection at LEP-200 its mass is always of the order of the
electroweak scale (and the dependence on the soft breaking scale is only
logarithmic). Similar bounds have been calculated for extended
supersymmetric models.
An analytical upper bound on the tree-level mass of the lightest Higgs
boson (LHB) is known for very general Supersymmetric Standard Models with
extended Higgs or gauge sectors \cite{eq,kane}. This bound depends on the
electroweak
scale (given by $M_Z$) and on the new Yukawa or gauge couplings that
appear in the theory. Numerical bounds can be obtained by placing
limits on these unknown parameters ({\it e.g.} assuming that the theory
remains perturbative up to some high scale).
These bounds are typically greater than the MSSM bound but still of
order $M_Z$.

In a recent paper \cite{pandita} a bound on the lightest Higgs boson mass
was calculated in a particular supersymmetric extended model with
spontaneous R-parity breaking, the supersymmetrized Singlet Majoron Model
(SSMM).
The obtained bound was found to be qualitatively different from the
general bounds of \cite{eq,kane}. In contrast with these general bounds
it was found a dependence on exotic vacuum expectation values (VEVs) that
are naturally of the order
of the SUSY breaking scale so that the bound is no longer controlled
by $M_Z$, although it turned out to be numerically
very similar to the MSSM bound due to the smallness of some
Yukawa couplings.
Theoretically this is a disturbing result, because it leaves open the
possibility of finding similar models in which the mass of the LHB
is much larger than $M_Z$ (evading the well behaved
general bounds of \cite{eq,kane}) with important consequences for the
Higgs phenomenology in such models.

The purpose of this letter is twofold. First of all, to re-analyze
the SSMM bound. In section 1 we will show how the
SSMM bound on the LHB tree-level mass can be improved,
eliminating all the dangerous dependence on exotic VEVs [this will
be true also after the inclusion of the most important one-loop radiative
corrections (from top-stop and bottom-sbottom loops), see section 2].
And second of all, to re-derive in detail the general bound
of Ref. \cite{eq} in order to clarify its applicability range. This will be
done in section 3.
\vspace{1cm}

{\bf 1.} The Supersymmetric Singlet Majoron Model \cite{gmpr} is the
simplest viable extension of the MSSM which can accommodate spontaneous
R-parity breaking (and so, lepton number breaking).
The extra fields added to the MSSM are right-handed
neutrino chiral superfields $N_i$ (with $L=-1$, and $i$ is a
family index, $i=1,2,3$) and a singlet superfield $\Phi$ (with $L=2$).
With this field content, the most general superpotential renormalizable
and gauge invariant is
\bear
\label{ssmmw}
f&=&\sum_{i,j}\corl h^u_{ij} Q_i\cdot H_2 U^c_j
+ h_{ij}^d H_1\cdot Q_i D_j^c +
h^{\nu}_{ij} L_i\cdot H_2 N_j \nonumber\\
&+& h_{ij}^e H_1\cdot L_i E_j^c
 - \lambda_{ij} N_i N_j \Phi \corr+ \mu H_1\cdot H_2,
\eear
where the notation is self-explanatory. The new Yukawa couplings
$h^{\nu}_{ij}$ are responsible for the mass of neutrinos and therefore
they have to be small. For the phenomenological restrictions used to
put bounds in these couplings see \cite{gmpr}.

The tree-level scalar potential for neutral states,
$V(H_1^0,H_2^0,\tilde{\nu_i}^0,{\tilde N}_j^0,\Phi^0)$ can be
readily derived (where the $\tilde{\nu_i}^0$ are the left-handed
sneutrinos). It consists of three parts: \be
V=V_F+V_D+V_{Soft},
\ee
with
\bear
\label{vfd}
V_F&=& \left|\mu H_2^0\right|^2 + \mod\mu H_1^0 +
\sum_{i,j}h^{\nu}_{ij}\tilde{\nu_{i}}^0 {\tilde N}_j^0\mod^2 +
\sum_i\mod\sum_j h^{\nu}_{ij}H_2^0{\tilde N}_j^0\mod^2\nonumber\\
&+&\sum_j\mod\sum_i( h^{\nu}_{ij}H_2^0\tilde{\nu_i}^0 - 2
\lambda_{ij}{\tilde N}_i^0\Phi^0)\mod^2
+\mod\sum_{i,j}\lambda_{ij}{\tilde N}_i^0{\tilde N}_j^0\mod^2,\vspace{.5cm}\\
V_D&=&\frac{1}{8}G^2 \left[\left|H_1^0\right|^2 - \left|H_2^0\right|^2 +
\sum_i\left|\tilde{\nu_i}^0\right|^2 \right]^2,
\eear
where $G^2\equiv g^2+g'^2$, and
\bear
\label{vsoft}
V_{Soft}&=&m_1^2 \left|H_1^0\right|^2 +m_2^2 \left|H_2^0\right|^2 +\sum_i
m_{{\tilde N}_i}^2
\left|{\tilde N}_i^0\right|^2 + m_{\phi}^2 \left|\Phi^0\right|^2 +\sum_i m_{
\tilde{\nu_i}}^2
\left|\tilde{\nu_i}^0\right|^2\nonumber\\
&-& \corl B\mu H_1^0 H_2^0 +
\sum_{i,j}A^{\nu}_{ij}h_{ij}^{\nu}H_2^0\tilde{\nu_i}^0{\tilde N}_j^0 -
\sum_{i,j}A^{\lambda}_{ij}\lambda_{ij}{\tilde N}_i^0 {\tilde N}_j^0\Phi^0
+ h.c. \corr .
\eear
For simplicity, we will neglect any possible CP breaking effects in
the following assuming that all the parameters in this potential are real.
Then we will take $\lambda_{ij}=\lambda_i\delta_{ij}$ by an appropriate
rotation of the $N_i$ fields.
As was shown in ref. \cite{gmpr}, in a wide region of parameter space,
not only $H_1^0$ and $H_2^0$ develop a VEV but also $\tilde{\nu_i}^0$,
${\tilde N}_j^0$ and $\Phi^0$ do:
\be
\langle H_1^0 \rangle = v_1,\;
\langle H_2^0 \rangle = v_2,\;
\langle \tilde{\nu_i}^0 \rangle = x_i,\;
\langle {\tilde N}_j^0 \rangle = y_j,\;
\langle \Phi \rangle = \phi,
\ee
with the hierarchy $x_i\sim h^{\nu} v\ll v\equiv\sqrt{v_1^2+v_2^2}
\ll y_j,\phi
\sim {\cal O} (M_S)$, ($M_S$
representing the scale of soft supersymmetry breaking, {\it e.g.}
$M_S\leq 1\ TeV$) thus leading to the spontaneous breaking of
$SU(2)_L\times U(1)_Y$, lepton number $L$ and R-parity. The
implications of this interesting scenario are further studied in
\cite{gmpr,more}.

Following the general procedure, an upper bound on
the tree level mass of the LHB in the SSMM can be
obtained studying the scalar mass matrix for the fields $H_1^{0r},
H_2^{0r},\tilde{\nu_i}^{0r},{\tilde N}_j^{0r},\Phi^{0r}$ [with
$\phi_i^0=(\phi_i^{0r}+i\phi_i^{0i})/\sqrt{2}$]. That matrix
is now a $9\times 9$
matrix but the analysis of the $2\times 2$ submatrix for $H_1^{0r},
H_2^{0r}$ leads easily to such a bound. This upper bound was calculated
in \cite{pandita} and is
\bear
\label{PB}
m_h^2&\leq &\frac{1}{2}G^2v^2\cos^22\beta + \sum_i x_i^2 \left[
\frac{m_{\tilde{\nu_i}}^2}{v^2} +\frac{1}{4}G^2\left(\cos 2\beta +\sum_j
\frac{x_j^2}{v^2}\right)\right] \nonumber\\
&+&\sum_{i,j,k,l}x_ix_jh^{\nu}_{ik}h^{\nu}_{jl}\left(\delta_{kl}\sin^2\beta
+\frac{y_k y_l}{v^2} \right),
\eear
with $v^2\equiv v_1^2+v_2^2$ and $\tan\beta\equiv v_2/v_1$.
As stressed in \cite{pandita}, there is an explicit dependence of the
bound on the exotic VEVs $x_i$ and $y_k$ and on the soft breaking mass of
sneutrinos, $m_{\tilde{\nu_i}}$. In particular, the bound (\ref{PB}) is
not finite in the formal limit $x_i,y_k,m_{\tilde{\nu_i}}\rightarrow\infty$,
that is, there is no decoupling of the exotic VEVs and soft masses
from the bound. The most important one
loop corrections to this tree level bound, coming from the top-stop and
bottom-sbottom sectors were also calculated in \cite{pandita} and it was
also found the same bad non-decoupling behaviour. Nevertheless, the
smallness of $h^{\nu}_{ij}$ and $x_i$ makes the bound
numerically indistinguishable from the MSSM bound $M_Z^2\cos^22\beta$.

Anyhow, as we are about to see, the bound (\ref{PB}) can in fact be improved,
that is, (\ref{PB}) is not saturated and a more stringent bound can be
found. Moreover, this new bound will turn out to be independent of
soft-breaking
masses or exotic VEVs and is always ${\cal O}(M_Z^2)$. In order to achieve
this improvement we need to examine larger mass submatrices. In fact, we will
need the $5\times 5$ mass matrix for the fields $H_1^{0r}$, $H_2^{0r}$ and
$\tilde{\nu_i}^{0r}$. The elements of this matrix,
derived from the effective potential (\ref{vfd}-\ref{vsoft}) are:
\bear
\label{mael}
M_{11}^2&=&\mu_1^2+\frac{1}{4}G^2(3v_1^2-v_2^2)+\frac{1}{4}G^2x^2
=-m_3^2\tan\beta+\frac{1}{2}G^2v_1^2-\mu\frac{\sigma^2}{v_1},\nonumber\\
M_{22}^2&=&\mu_2^2-\frac{1}{4}G^2(v_1^2-3v_2^2)-\frac{1}{4}G^2x^2
+\sum_i\parl \sum_j h^{\nu}_{ij} y_j \parr^2
+\sum_j\parl \sum_i h^{\nu}_{ij} x_i \parr^2\nonumber\\
&=&-m_3^2\cot\beta+\frac{1}{2}G^2v_2^2+\frac{R^3}{v_2},\nonumber\\
M_{12}^2&=&m_3^2-\frac{1}{2}G^2v_1v_2,
\eear
with $\mu_i^2=|\mu|^2+m_i^2$, $x^2=\sum_i x_i^2$ and $m_3^2=-B\mu$.
The mass parameters $\mu_1$ and $\mu_2$ have been traded by the VEVs
$v_1$ and
$v_2$ using the minimization conditions $\partial V/\partial v_1=
\partial V/\partial v_2=0$:
\bear
\label{minim}
\mu_1^2&=&-\frac{1}{4}G^2v^2\cos 2\beta -m_3^2\tan\beta-\mu
\frac{\sigma^2}{v_1}-\frac{1}{4}G^2x^2,\\
\mu_2^2&=&\frac{1}{4}G^2v^2\cos 2\beta -m_3^2\cot\beta+
\frac{R^3}{v_2}+\frac{1}{4}G^2x^2
-\sum_i\parl \sum_j h^{\nu}_{ij} y_j \parr^2
-\sum_j\parl \sum_i h^{\nu}_{ij} x_i \parr^2,\nonumber
\eear
where
\bear
\label{defi}
\sigma^2&\equiv&\sum_{i,j}h^{\nu}_{ij}x_iy_j,\nonumber\\
R^3&\equiv&\sum_{i,j}h^{\nu}_{ij}(2\lambda_j\phi+A^{\nu}_{ij})x_iy_j.
\eear

Due to the
breaking of lepton number, Higgses and sneutrinos mix and the corresponding
matrix elements are non vanishing:
\bear
\label{maelmix}
M_{\tilde{\nu_i}1}^2&=&\mu\sigma_i+\frac{1}{2}G^2v_1x_i,\nonumber\\
M_{\tilde{\nu_i}2}^2&=&-\sum_j h^{\nu}_{ij}y_j(2\lambda_j\phi+A^{\nu}_{ij})
-\frac{1}{2}G^2v_2x_i +
2v_2\sum_{j,k}h^{\nu}_{ij}h^{\nu}_{kj}x_k,
\eear
where $\sigma_i\equiv\sum_jh^{\nu}_{ij}y_j$. And finally we also need the
sneutrino mass matrix
\bear
\label{maelsne}
M_{\tilde{\nu_i}\tilde{\nu_j}}^2
&=&\parl m_{\tilde{\nu_i}}^2+\frac{1}{4}G^2x^2+\frac{1}{4}G^2v^2\cos
2\beta\parr \delta_{ij}+\sigma_i\sigma_j+\frac{1}{2}G^2x_ix_j+v_2^2\sum_k
h^{\nu}_{ik}h^{\nu}_{jk}.
\eear
For later use we also obtain from the condition $\partial V/\partial x_i=0$
the relation
\be
\label{minix}
\sum_i m_{\tilde{\nu_i}}^2
x_i^2 = v_2 R^3 -\frac{1}{4}G^2x^2(x^2+v^2\cos 2\beta)
-\mu v_1\sigma^2 -\sigma^4 -v_2^2\sum_j \parl\sum_i h^{\nu}_{ij}x_i\parr^2,
\ee
from which sneutrino masses can be traded by other parameters of the
potential.

Next we define the normalized field $\tilde{\nu}^{0}$ as
\be
\tilde{\nu}^{0}\equiv \frac{\sum_ix_i\tilde{\nu_i}^{0}}{\sqrt{\sum_ix_i^2}}=
\frac{1}{x}\sum_ix_i\tilde{\nu_i}^{0},
\ee
such that $\langle\tilde{\nu}^{0}\rangle=x$, while any combination of the
fields $\tilde{\nu_i}^{0}$ orthogonal to $\tilde{\nu}^{0}$ has a
vanishing VEV. We also define the new field $H'_1{}^{0}$ as the combination
\be
\label{rot}
H'_1{}^{0}=\frac{v_1H_1^{0}+x\tilde{\nu}^{0}}{\sqrt{v_1^2+x^2}}\equiv
\frac{1}{v'_1}(v_1 H_1^{0}+x\tilde{\nu}^{0}),
\ee
so that $\langle H'_1{}^{0}\rangle=v'_1\equiv\sqrt{v_1^2+x^2}$. This
redefinition of fields amounts to a change of basis from $(H_1^{0r},
H_2^{0r},\tilde{\nu_i}^{0r})$ to $(H'_1{}^{0r},H_2^{0r},...)$ where
now the only doublet fields having a non zero VEV are $H'_1{}^{0r}$ and
$H_2^{0r}$.
The important point is that, in this rotated basis, the $2\times 2$
mass submatrix (for $H'_1{}^{0r},H_2^{0r}$), $M'_{ij}{}^2$, has a simpler form.
Using (\ref{minix}),
\bear
M'_{11}{}^2&=&\frac{1}{v'_1{}^2}\corl v_1^2M_{11}^2 + 2 v_1 \sum_i x_i
M_{\tilde{\nu_i}1}^2 + \sum_{i,j}x_ix_jM_{\tilde{\nu_i}\tilde{\nu_j}}^2
\corr\nonumber\\
&=&\frac{1}{v'_1{}^2}\left[-v_1^2m_3^2\tan\beta+\frac{1}{2}G^2v'_1{}^4+
R^3v_2\right],\\
M'_{22}{}^2&=&M_{22}^2=-m_3^2\cot\beta+\frac{1}{2}G^2v_2^2 +\frac{R^3}{v_2},\\
M'_{12}{}^2&=&\frac{1}{v'_1}\corl v_1 M_{12}^2+\sum_i x_i
M_{\tilde{\nu_i}2}^2\corr
\nonumber\\
&=&\frac{1}{v'_1}\corl m_3^2v_1-\frac{1}{2}G^2v'_1{}^2v_2-R^3
+2v_2\sum_j\parl\sum_ih^{\nu}_{ij}x_i\parr^2
\corr.
\eear
Or, writing
\be
m'_3{}^2\equiv\frac{1}{v'_1}(m_3^2 v_1-R^3);\;\;\;\;\;
h'_j{}^2\equiv\frac{1}{v'_1}\sum_ih^{\nu}_{ij}x_i;\;\;\;\;\;\;
\tan\beta'\equiv\frac{v_2}{v'_1},
\ee
(note that this last definition is the natural one in this model) the
$2\times2$ submatrix takes the form:
\begin{eqnarray}
 M^2&=&\left|\left|
\begin{array}{cc}
-m'_{3}{}^2\tan\beta'+\frac{1}{2}G^2 v'_1{}^2&
m'_{3}{}^2 - \frac{1}{2}G^2 v'_1 v_2 + 2\sum_i h'_i{}^2 v'_1 v_2\\
&\\
m'_{3}{}^2 - \frac{1}{2}G^2 v'_1 v_2 + 2\sum_i h'_i{}^2 v'_1 v_2&
-m'_{3}{}^2\cot\beta'+ \frac{1}{2}G^2 v_2^2 \\
\end{array}\right|\right|.
\end{eqnarray}
As is well known, the full $9\times 9$ scalar mass matrix must have an
eigenvalue
smaller than (or equal to) the smallest eigenvalue of this $2\times 2$
submatrix. In
this way the following upper limit for the LHB mass results:
\be
\label{ssmmb}
m_h^2\leq
\frac{1}{2}G^2v'{}^2\cos^22\beta'+v'{}^2\sum_ih'{}^2_i\sin^22\beta',
\ee
where $v'^2\equiv v_1^2+v_2^2+x^2$. Note that now
\be
M_Z^2=\frac{1}{2}G^2v'{}^2=\frac{1}{2}(g^2+g'^2)
\corl v_1^2+v_2^2+\sum_ix_i^2\corr,
\ee
[$\langle\tilde{\nu_i}^0\rangle=x_i$ breaks $SU(2)_L\times U(1)_Y$ and then
contributes to the gauge boson masses] so that $v'$ is fixed to be $174\ GeV$.

Remarkably, the bound (\ref{ssmmb}) has the same form as the general
bound calculated for supersymmetric models with an extended Higgs sector
\cite{eq} and so it
exhibits its same good properties, namely to be controlled exclusively by
the electroweak scale (note that the rotated Yukawa couplings $h'_j=\sum_i
h^{\nu}_{ij}x_i/v'_1$ are well behaved even in the formal limit $x_i\rightarrow
\infty$ because $x_i/v'_1<1$).
\vspace{1cm}

{\bf 2.} It is natural to ask whether the good effect of this field rotation
also extends to radiative corrections. That is, do the one-loop
radiative corrections to the bound (\ref{ssmmb}) exhibit decoupling
of the exotic VEVs $x_i, y_k$  and soft-breaking masses $m_{\tilde{\nu_i}}$?
We will show in this section that
this is actually what happens for the most important radiative corrections:
the ones coming from the top-stop and bottom-sbottom sectors.

The leading terms of the one-loop corrections to the Higgs mass can be
easily calculated using the well known expression for the one-loop
contribution to the effective potential (in $\overline{DR}$ scheme)
\be
\label{dvo}
\Delta V_1 = \frac{1}{64\pi^2} Str M_i^4\left(\log\frac{M_i^2}{Q^2}
-\frac{3}{2}\right),
\ee
where $Q$ is the renormalization scale and $M_i$ are the field-dependent
masses of the different species of particles.
To fix the notation we list here the relevant masses for
the top-stop sector, which are given by
\be
m_t^2(v_2)=h_t^2v_2^2,
\ee
\begin{eqnarray}
\label{stopm}
 M_{\tilde t}^2(v_1,v_2,x_i,y_j)&=&\left|\left|
\begin{array}{cc}
m_{\tilde Q}^2+ m_t^2 +M_Z^2 D^t_{L} \cos 2\beta &
h_t(A_t v_2 + \mu v_1 + \sigma^2)\\
&\\
h_t(A_t v_2 + \mu v_1 + \sigma^2) &
m_{\tilde U}^2+ m_t^2 +M_Z^2 D^t_{R} \cos 2\beta
\end{array}\right|\right|,
\end{eqnarray}
with $m_{\tilde Q}^2,m_{\tilde U}^2$ the soft masses for left and
right-handed stops, $A_t$ the
trilinear coupling associated with the term $h_t Q_3\cdot H_2 U_3^c$
in the superpotential and
\be
D^t_{L}=\frac{1}{2}-\frac{2}{3}\sin^2\theta_W,\;\;\;\;\;\;
D_{R}^t=\frac{2}{3}\sin^2\theta_W,
\ee
while $\sigma^2$ was defined in (\ref{defi}). We will call $\mtu,\mtd$
the two eigenvalues of (\ref{stopm}) with $\mtu\geq\mtd$.

For the bottom-sbottom sector we have
\be
m_b^2(v_1)=h_b^2v_1^2,
\ee
\begin{eqnarray}
\label{sbotm}
 M_{\tilde b}^2(v_1,v_2)&=&\left|\left|
\begin{array}{cc}
m_{\tilde Q}^2+ m_b^2 +M_Z^2 D^b_L \cos 2\beta &
h_b(A_b v_1 + \mu v_2 )\\
&\\
h_b(A_b v_1 + \mu v_2 ) &
m_{\tilde D}^2+ m_b^2 +M_Z^2 D^b_R \cos 2\beta
\end{array}\right|\right|,
\end{eqnarray}
with $m_{\tilde D}^2$ the soft mass for right-handed sbottoms, $A_b$ the
trilinear coupling associated with the term $h_b H_1\cdot Q_3 D_3^c$ in the
superpotential and
\be
D^b_L=-\frac{1}{2}+\frac{1}{3}\sin^2\theta_W,\;\;\;\;\;\;
D^b_R=-\frac{1}{3}\sin^2\theta_W.
\ee
We will call $\mbu,\mbd$
the two eigenvalues of (\ref{sbotm}) with $\mbu\geq\mbd$.
As we are not considering the gauge contributions to (\ref{dvo}), we
will also neglect the $D$ term contributions to (\ref{stopm}) and
(\ref{sbotm}) in the following.

The one-loop corrected effective potential is a function of
$\phi_i=(H_1^{0r},H_2^{0r},\tilde{\nu}_i^{0r},{\tilde N}_j^{0r},
\phi^{0r})$ [or
equivalently $(v_1,v_2,x_i,y_j,\phi)$] and
so, the squared-mass matrix $M_{ij}^2$ will receive a one-loop
correction $\partial^2\Delta V_1/\partial\phi_i\partial\phi_j$. After
correcting also the minimization conditions (\ref{minim}) by
including the one-loop contribution $\partial\Delta V_1/\partial\phi_i$
we obtain
\bear
M^2_{11}{}^{(1)}&=&-(m_3^2+\delta m_3^2)\tan\beta +\frac{1}{2}G^2v_1^2
-\frac{1}{v_1}\mu\sigma^2(1+\delta f)+\Delta_{11},\nonumber\\
M^2_{22}{}^{(1)}&=&-(m_3^2+\delta m_3^2)\cot\beta +\frac{1}{2}G^2v_2^2
+\frac{R^3}{v_2}-\frac{A_t}{v_2}\sigma^2\delta f+\Delta_{22},\nonumber\\
M^2_{12}{}^{(1)}&=&(m_3^2+\delta m_3^2) - \frac{1}{2}G^2v_1v_2+\Delta_{12},
\eear
with
\bear
\delta m_3^2&=&\frac{3}{32\pi^2}\left\{\frac{h_t^2A_t\mu}{m_{\tilde{t_1}}^2
-m_{\tilde{t_2}}^2}\left[f(\mtu )-f(\mtd )
\right]+(t\rightarrow b)
\right\},\nonumber\\
\delta f&=&\frac{3h_t^2}{32\pi^2}\frac{f(\mtu )-f(\mtd
)}{\mtu-\mtd},\nonumber\\
\Delta_{11}&=&\frac{3}{8\pi^2}\llavl h_t^2m_t^2\mu^2A'_T{}^2g(\mtu,\mtd)
\nonumber\\
&+&h_b^2m_b^2\left.\left[\log\frac{\mbu\mbd}{m_b^4}
+2A_bA_B\log\frac{\mbu}{\mbd}
+A_b^2A_B^2g(\mbu,\mbd)
\right]
\right\},\nonumber\\
\Delta_{22}&=&\frac{3}{8\pi^2}\llavl h_b^2m_b^2\mu^2A_B^2g(\mbu,\mbd)
\nonumber\\
&+&h_t^2m_t^2\left.\left[\log\frac{\mtu\mtd}{m_t^4}
+2A_tA'_T\log\frac{\mtu}{\mtd}
+A_t^2A'_T{}^2g(\mtu,\mtd)
\right]
\right\},\nonumber\\
\Delta_{12}&=&\frac{3}{8\pi^2}\left\{
h_t^2m_t^2\mu A'_T\left[\log\frac{\mtu}{\mtd}
+A_tA'_Tg(\mtu,\mtd)
\right]
\right.\nonumber\\
&+&\left. h_b^2m_b^2\mu A_B\left[\log\frac{\mbu}{\mbd}
+A_bA_Bg(\mbu,\mbd)
\right]\right\},\nonumber\\
\eear
and
\bear
f(m^2)=2m^2\left[\log\frac{m^2}{Q^2}
-1
\right],\;\;\;\;\;\;
g(m_1^2,m_2^2)=2-\frac{m_1^2+m_2^2}{m_1^2-m_2^2}\log
\frac{m_1^2}{m_2^2},
\eear
\bear
A'_T=\frac{1}{\mtu-\mtd}\left(A_t+\mu\cot\beta+\frac{\sigma^2}{v_2}
\right),\;\;\;\;\;\;
A_B=\frac{A_b+\mu\tan\beta}{\mbu-\mbd}.
\eear
All of the matrix elements $M_{ij}^{2}{}^{(1)}$ $(i,j=1,2)$ diverge
in the decoupling SUSY limit $\mu y_k, m_3^2+\delta m_3^2\rightarrow\infty$.
In fact, writing $\mu^2\sim y_k^2\sim m_3^2+\delta m_3^2\sim {\cal O}
(M_S^2)$, all the  $M_{ij}^{2}{}^{(1)}$ grow like $M_S^2$:
\bear
M^2_{11}{}^{(1)}&=&-(m_3^2+\delta m_3^2)\tan\beta
-\frac{1}{v_1}\mu\sigma^2(1+\delta f)+...,\nonumber\\
M^2_{22}{}^{(1)}&=&-(m_3^2+\delta m_3^2)\cot\beta
+\frac{R^3}{v_2}-\frac{A_t}{v_2}\sigma^2\delta f+...,\nonumber\\
M^2_{12}{}^{(1)}&=&(m_3^2+\delta m_3^2) +...,
\eear
where the dots stand for contributions that are finite in the decoupling limit.
The eigenvalues of this matrix will grow also like $M_S^2$ if $Det M^2{}^{(1)}$
grows like $M_S^4$, but, if due to some cancellation, $Det M^2{}^{(1)}$
grows only like $M_S^2$ one of the eigenvalues (the lightest) will remain
finite in the decoupling limit \cite{comelli}.
In fact one can see that this cancellation
takes place for the $(m_3^2+\delta m_3^2)^2$ contribution to $Det M^2{}^{(1)}$
but not for the rest of the terms. Then, as was found in \cite{pandita},
the bound derived from this unrotated submatrix receives one-loop
corrections which are not controlled only by the electroweak scale.

To study the one-loop rotated matrix we need also
\bear
M^2_{\tilde{\nu_i}1}{}^{(1)}&=&M^2_{\tilde{\nu_i}1}
+\mu\sigma_i\eta,\nonumber\\
M^2_{\tilde{\nu_i}2}{}^{(1)}&=&M^2_{\tilde{\nu_i}2}+A_t\sigma_i\eta
+\frac{3}{8\pi^2}h_t^2m_t^2A'_T\sigma_i\log\frac{\mtu}{\mtd},\nonumber\\
M^2_{\tilde{\nu_i}\tilde{\nu_j}}{}^{(1)}&=&
M^2_{\tilde{\nu_i}\tilde{\nu_j}}+
\sigma_i\sigma_j\eta,
\eear
with the tree level matrix elements on the right hand side as given in
(\ref{maelmix}) and (\ref{maelsne}) and
\be
\eta=\frac{3h_t^2}{8\pi^2}
m_t^2 A'_T{}^2g(\mtu,\mtd)+\delta f.
\ee

Note that $\sum_{i,j}M_{\tilde{\nu_i}\tilde{\nu_j}}^2x_ix_j$
contains the term
$\sum_im_{\tilde{\nu_i}}^2x_i^2$ which receives also the one-loop correction
\be
\delta \sum_im_{\tilde{\nu_i}}^2x_i^2 =-\frac{1}{2}\sum_ix_i
\frac{\partial\Delta V_1}{\partial x_i}=-\frac{3h_t^2}{32\pi^2}v_2
A'_T\sigma^2\left[ f(\mtu)-f(\mtd)
\right].
\ee
The rotated $2\times 2$ mass matrix then takes the one-loop form:
\begin{eqnarray}
\label{rotcorr}
 M'{}^2&=&\left|\left|
\begin{array}{cc}
-(m'_3{}^2+\delta m'_3{}^2)\tan\beta'+ \Delta'_{11}&
(m'_3{}^2+\delta m'_3{}^2) + \Delta'_{12}\\
&\\
(m'_3{}^2+\delta m'_3{}^2) + \Delta'_{12}&
-(m'_3{}^2+\delta m'_3{}^2)\cot\beta'
+ \Delta'_{22}
 \\
\end{array}\right|\right|.
\end{eqnarray}
with
\be
\delta m'_3{}^2=\frac{1}{v'_1}\corl \delta m_3{}^2
v_1  + A_t\sigma^2\delta f \corr.
\ee
The terms $\Delta'_{ij}$ are finite in the decoupling limit and are
given by:
\bear
\Delta'_{11}&=&\frac{1}{2}G^2v'_1{}^2+\frac{3}{8\pi^2}h_t^2m_t^2A'_T{}^2
\frac{\sigma^2}{v'_1{}^2}
(\sigma^2+2\mu v_1)g(\mtu,\mtd)+\frac{v_1^2}{v'_1{}^2}\Delta_{11},
\nonumber\\
\Delta'_{22}&=&\frac{1}{2}G^2v_2^2+\Delta_{22},\nonumber\\
\Delta'_{12}&=&-\frac{1}{2}G^2v'_1v_2+2\sum_j h'_j{}^2v'_1v_2
+\frac{3}{8\pi^2}\frac{1}{v'_1}h_t^2m_t^2\sigma^2\eta A_tA'_T{}^2g(\mtu,\mtd)
\nonumber\\
&+&\frac{3}{8\pi^2}\frac{1}{v'_1}h_t^2m_t^2\sigma^2A'_T\log
\frac{\mtu}{\mtd}
+\frac{v_1}{v'_1}\Delta_{12}.
\eear
A look at (\ref{rotcorr}) shows immediately that there is decoupling, that
is, $Det M'{}^2=M'_{11}{}^2M'_{22}{}^2-M'_{12}{}^4\sim{\cal O}(M_S^2)$.

In fact, the one-loop corrected version of the tree-level bound (\ref{ssmmb})
takes the simple form
\bear
m_h^2&\leq &\Delta'_{11}\cos^2\beta'+\Delta'_{22}\sin^2\beta'+
\Delta'_{12}\sin2\beta'
=M_Z^2\cos^2 2\beta' + 2 M_W^2 \sum_i \frac{h'_i{}^2}{g^2}\sin^2
2\beta'\nonumber\\
&+&\frac{3g^2}{16\pi^2}\left\{\frac{m_t^4}{m_W^2}\left[
\log\frac{\mtu\mtd}{m_t^4}+
\tilde{X}^2_t\left(2\log\frac{\mtu}{\mtd} + \tilde{X}^2_t
g(\mtu,\mtd)\right)\right]+ (t\rightarrow b)\right\},
\eear
where
\be
\tilde{X}_t^2=\frac{(A_t+\mu\cot\beta+\sigma^2/v_2)^2}{\mtu-\mtd};
\;\;\;\;
\tilde{X}_b^2=\frac{(A_b+\mu\tan\beta)^2}{\mbu-\mbd}.
\ee
Note that the one-loop corrections have exactly the same form as in the
MSSM, the only difference arising in $\tilde{X}_t^2$ which now includes
the term $\sigma^2/v_2$.
\vspace{1cm}

{\bf 3.} In this section we will re-derive the bound on the lightest Higgs
boson mass for general Supersymmetric Standard Models under
very general assumptions. The particular form of the bound for
some special cases of interest \cite{eq} will be presented, and
at the end we will show how to apply the general bound to the SSMM case
finding the same result obtained in the direct calculation of section 1.

Let us first assume that the Higgs sector of the general Supersymmetric
model under consideration contains at least two $SU(2)_L$ doublets
($H_1$, $H_2$ with hypercharges $\pm 1/2$) taking vacuum
expectation values
\be
\begin{array}{cc}
\left.
\begin{array}{c}
\langle H_1\rangle=v_1\\
\langle H_2\rangle=v_2
\end{array}\right\}&v^2\equiv v_1^2+v_2^2\leq (174\ GeV)^2,
\end{array}
\ee
with the equality holding when only these two doublets drive electroweak
breaking. In the general case other fields may contribute to gauge boson
masses. When there are extra doublets ($d$ doublets with hypercharge
$-1/2$ and $d$ with hypercharge $1/2$) is well known that a field rotation
can be made such that only one doublet of each type takes a non zero VEV
\cite{rot} (that can be taken real and positive if electric charge is
conserved). In that case these two rotated fields will be the ones we
are calling $H_1$ and $H_2$ (as we did for the SSMM in the previous
sections). In general other fields in higher $SU(2)_L$ representations
can participate in the electroweak breaking (but satisfying the constraints
from $\Delta\rho$). In the following we will make the reasonable
assumption that no fields in representations
higher than triplets take non-zero VEVs\footnote{The vast majority of
the models of interest satisfy this
requirement although a bound that applies even without this restriction
can be derived (see Ref. \cite{sca}).} (models with unsuppressed triplet
VEVs have been studied in the literature).

In general the LHB has the same quantum numbers as
the Standard Model Higgs so that we will concentrate in the study
of the CP even Higgs sector\footnote{This terminology is only
valid when CP is a good symmetry (as in the MSSM), but we do not
need this assumption for the derivation to be correct.}.
Let us consider the most general tree-level potential for the fields
$H_1^{0r},H_2^{0r}$ (coming from a renormalizable full scalar potential):
\begin{eqnarray}
\label{vtree}
V&=&V_0+M_1^3 \Hu + M_2^3 \Hd\nonumber\\
&+& m_1^2 (\Hu)^2 + 2 m^2_{12}\Hu\Hd + m^2_2 (\Hd)^2\nonumber\\
&+& M_{11} (\Hu)^3 + M_{12} (\Hu)^2\Hd +M_{21} (\Hd)^2\Hu
+ M_{22} (\Hd)^3 \nonumber\\
&+&\lambda_1 (\Hu)^4 +\lambda_{12} (\Hu)^2(\Hd)^2 +\lambda_2
(\Hd)^4\nonumber\\
&+& \lambda'_{12} (\Hu)^3 \Hd +  \lambda'_{21} \Hu(\Hd)^3,
\end{eqnarray}
In general, the parameters appearing in this potential are some
function of other scalar fields, {\em e.g.}
$M_1\equiv M_1(\chi^{0r},\xi^{0r},...)$.
The superindex $o$ will indicate that such functions have to be evaluated
with these extra scalar fields at their vacuum expectation values:
$M_1^o\equiv M_1(\chi^{0r}=
\langle\chi^{0r}\rangle,\xi^{0r}=\langle\xi^{0r}\rangle,...)$.

By symmetry considerations some of the terms in this potential can be
forbidden. For example, imposing invariance of the potential under the
symmetry transformation $H_i^{0r}\rightarrow -H_i^{0r}$ many couplings
in (\ref{vtree}) should be set to zero. As we will see later, we do not
need to impose this symmetry by hand but it will arise automatically for
the quartic couplings of the supersymmetric theories in which we are
interested.

As $\Hu, \Hd$ are the neutral components (more precisely the real part)
of two $SU(2)_L$ doublets and the full potential has to be gauge
invariant, it can be immediately deduced that $M^3_{1,2}$ must come from a
field or combination of fields transforming as a $SU(2)_L$ doublet,
that is
\be
\label{Mi3}
(M^o_{1,2})^3\sim m^2 \langle d \rangle,
\ee
with $m^2$ some gauge invariant squared mass and $\langle d \rangle$
representing the VEV of some $SU(2)_L$ doublet (fundamental or not,
but different from $\Hu$ and $\Hd$ by construction). Now, as
$H_1$ and $H_2$ are the only doublets with a non vanishing VEV (and no
fields in representations higher than triplets take a non-zero VEV) we get
the restriction
\be
\label{Mi0}
(M^o_{1})^3 = (M^o_{2})^3 = 0.
\ee
Using similar arguments, $SU(2)_L\times U(1)_Y$ invariance of the
full potential implies\footnote{By assumption we discard the possibility of
$M^o_{ij}$ transforming as a $SU(2)$ quadruplet and taking a VEV.}
\be
\label{Mij}
M^o_{ij}\sim \langle d \rangle,
\ee
so that one can also deduce the relations
\be
\label{Mij0}
M^o_{11} = M^o_{12} =M^o_{21} = M^o_{22} = 0.
\ee

Next, taking into account that the scalar potentials
we are considering come from a (softly broken) supersymmetric theory,
we will get some restrictions on the quartic couplings in $V$.
As is well known, these quartic couplings are of fundamental importance
for the tree-level upper bound we want to derive.
Let us consider separately the two different types of supersymmetric
contributions to the effective potential.

$i)$ $F$ terms

These are given by the well known formula
\be
\label{vf}
V_F=\sum_i\left|\frac{\partial f}{\partial \phi_i}\right|^2,
\ee
with $f(\phi_i)=W(\hat{\phi_i}\rightarrow \phi_i)$ [
$W(\hat{\phi_i})$ is the superpotential,
$\hat{\phi_i}$ the chiral superfields and
$\phi_i$ their scalar components].
The superpotential $W$ is at most cubic in the superfields
implying that the quartic F-terms in the potential must come
from the cubic terms in $W$. It is easy to see that quartic terms like
$(\Hu)^3\Hd$ or $(\Hd)^3\Hu$ can only arise if both a superfield
$\hat{\phi_i}$ and its hermitian conjugate $\hat{\phi_i}^*$ appear in
$W$, which is not possible ($W$ being an analytic function of the
chiral superfields). This proves that $F$-terms do not contribute to
the $\lambda'_{ij}$ couplings.

$ii)$ $D$ terms

The contribution of $D$ terms to the scalar potential takes the form
\be
\label{vd}
V_D=\frac{1}{2}\sum_a g_a^2 \left|\phi_iT_{ij}^a\phi_j^*\right|^2,
\ee
where $a$ runs over the gauge groups (with coupling constant $g_a$) and
$T^a$ are the generating matrices of group $a$  in the representation
of the fields $\phi_i$. According to (\ref{vd}) the only quartic terms
obtained are of the form
$H_1 H_1 H_1^* H_1^*$, $H_1 H_2 H_1^* H_2^*$ or $H_2 H_2 H_2^* H_2^*$.
So, neither $D$ terms  nor $F$ terms contribute to the
$\lambda'$ couplings and then
\be
\label{lp}
\lambda'_{12}=\lambda'_{21}=0.
\ee

Inserting eqs. (\ref{Mi0}), (\ref{Mij0}) and (\ref{lp})
in (\ref{vtree}), the tree-level scalar potential for $\Hu$, $\Hd$
takes the following form
\begin{eqnarray}
\label{vstree}
V_S(\Hu,\Hd)&=&V_0^o+\frac{1}{2}(m^o_1)^2 (\Hu)^2 + (m^o_{12})^2\Hu\Hd
+ \frac{1}{2}(m^o_2)^2 (\Hd)^2\nonumber\\
&+&\frac{1}{4}\lambda_{11} (\Hu)^4 +\frac{1}{4}\lambda_{12}
(\Hu)^2(\Hd)^2 +\frac{1}{4}\lambda_{22}
(\Hd)^4.
\end{eqnarray}
After writing $H_i^{0r}=h_i^{0r}+\sqrt{2}v_i$ [which corresponds to
$\langle H_i^0\rangle\equiv \langle(H_i^{0r}+iH_i^{0i})/2\rangle=v_i$],
$m_1^o$ and $m_2^o$ can be expressed as functions of the other parameters of
the potential just imposing that $V_S$ has its minimum at $(v_1,v_2)$ ,
that is, using the minimization conditions
\be
\label{mincon}
\frac{\partial V_S}{\partial h_i^{0r}}=0,\:\:\: (i=1,2).
\ee
In that way we are led to
\begin{eqnarray}
\label{m1m2}
m_1^2&=&-m_{12}^2\tan\beta-2\lambda_{11}v_1^2-\lambda_{12}v_2^2,\nonumber\\
m_2^2&=&-m_{12}^2\cot\beta-2\lambda_{22}v_2^2-\lambda_{12}v_1^2,
\end{eqnarray}
where $\tan\beta=v_2/v_1$ and the superindex $o$ is omitted for simplicity.

The analysis of the mass submatrix ($\cal{M}$$^2$) for the fields $\hu,\hd$
will give us the mass bound on the LHB mass.
This submatrix is now:
\begin{eqnarray}
\label{subm}
{\cal M}^2&=&\left|\left|
\begin{array}{cc}
m_1^2+6\lambda_{11} v_1^2+\lambda_{12} v_2^2 &
m_{12}^2+2\lambda_{12} v_1 v_2\\
&\\
m_{12}^2+2\lambda_{12} v_1 v_2 &
m_2^2+6\lambda_{22} v_2^2+\lambda_{12} v_1^2
\end{array}\right|\right|\nonumber\\
&&\nonumber\\
&=&\hspace{.5cm}\left|\left|
\begin{array}{cc}
-m_{12}^2\tan\beta+4\lambda_{11} v_1^2&
m_{12}^2+2\lambda_{12} v_1 v_2\\
&\\
m_{12}^2+2\lambda_{12} v_1 v_2 &
-m_{12}^2\cot\beta+4\lambda_{22} v_2^2 \\
\end{array}\right|\right|,
\end{eqnarray}
where the minimization conditions (\ref{m1m2}) have been used
to write the second expression. It is simple to show that the eigenvalues
of this matrix can be written as
\begin{eqnarray}
\label{mpm}
m_{\pm}^2&=&\left.\frac{1}{2}\right\{ -\overline{m}_{12}^2 +
4\lambda_{11}v_1^2 + 4\lambda_{22} v_2^2 \pm
\left[\left(\overline{m}_{12}^2+4 v_{11}^2 -4
v_{22}^2\right)^2\cos^22\beta
\right.\nonumber\\
&+& \left.\left.\left(\overline{m}_{12}^2+2\lambda_{12} v^2
\right)^2\sin^22\beta\right]^{1/2} \right\},
\end{eqnarray}
with $\overline{m}_{12}^2\equiv 2m_{12}^2/\sin2\beta$ and
$v_{ii}^2\equiv \lambda_{ii}v_i^2/\cos2\beta$. Using the inequality
\be
\label{desig}
\left[a^2\cos^22\beta+b^2\sin^22\beta\right]^{1/2}\geq a \cos^22\beta
+ b\sin^22\beta,
\ee
the following bound results:
\be
\label{central}
m_h^2\leq m_-^2\leq \left(4\lambda_{11}\cos^4\beta +
4\lambda_{22}\sin^4\beta +\lambda_{12}\sin^22\beta\right)v^2,
\ee
which is the central formula we were looking for.

Note that the bound is determined by the quartic couplings, as anticipated.
It implies that the bound is not sensitive to the details of the soft-breaking.
Then the only scale that enters the bound is $v$, the electroweak scale
and so, even if the soft breaking scale gets large there is always a light
scalar Higgs [that is, of mass
${\cal O}(M_Z)$] in the spectrum. One can go beyond (\ref{central}) and
obtain the particular form of the bound in some classes of models:

{\bf 3.1 Models with an extended Higgs sector.}

Having obtained the general formula (\ref{central}), we can now find
the particular form of the Higgs mass bound in a class of
nonminimal Supersymmetric models in which the Higgs sector is extended and
contains, apart from the two MSSM Higgs doublets, $H_1,\: H_2$,
other extra fields (singlets, more doublets without VEVs, triplets, etc.
See refs. [11-15]).

In that class of models the $\lambda_{ij}$ couplings in (\ref{central})
come from:

$i)$ $F$ terms

The only cubic terms in the superpotential that can give a contribution
are of the form
\be
\label{cubi}
\Delta W \sim \lambda \hat{\phi}\hat{H_i}\hat{H_j},
\ee
with $i,j=1,2$, and $\hat{\phi}$ an extra chiral superfield (note that
there are no such terms in the MSSM).
More specifically with a superpotential containing a part like
\be
\label{fh}
f=\lambda_0\phi_0 H_1^0 H_2^0 + \frac{1}{2}\lambda_1\phi_1 H_1^0
H_1^0 +\frac{1}{2}\lambda_{-1}\phi_{-1} H_2^0
H_2^0+\ldots,
\ee
the $\lambda_{ij}$ couplings in (\ref{vstree}) receive the contributions
\be
\label{deltalh}
\delta\lambda_{11}=\frac{1}{4}\lambda_1^2\:,\:\:\:
\delta\lambda_{22}=\frac{1}{4}\lambda_{-1}^2\:,\:\:\:
\delta\lambda_{12}=\lambda_0^2\:.
\ee
On top of that, note that $SU(2)_L\times U(1)_Y$ invariance
of the superpotential requires the $\phi_k$ scalars to be singlets or
neutral components of triplets with hypercharge $Y=0,\pm 1$.

$ii)$ $D$ terms

The contributions are exactly the same as in the MSSM:
\begin{eqnarray}
\label{lmssm}
\delta\lambda_{11}=\delta\lambda_{22}=\frac{1}{8}
\left(g^2+g'^2\right),\nonumber\\
\delta\lambda_{12}=-\frac{1}{4}\left(g^2+g'^2\right),
\end{eqnarray}
so that the bound on the LHB in these non-minimal models
is
\be
\label{cotah}
\frac{m_h^2}{v^2}\leq \frac{1}{2}\left(g^2+g'^2\right)\cos^22\beta
+\lambda_{0}^2\sin^22\beta+
\lambda_{1}^2\cos^4\beta +
\lambda_{-1}^2\sin^4\beta ,
\ee
with $v^2\equiv v_1^2+v_2^2$ as usual.
This bound was first obtained in \cite{eq}.

Of course, if $\lambda_k=0$ the MSSM bound is recovered. The non-minimal
correction, dependent on the new Yukawa couplings, $\lambda_k$,
is positive definite so that the bound is weaker than in the MSSM.
Also note that it is necessary to write the bound with an explicit $v^2$
dependence instead of using the formula $(g^2+g'^2)v^2/2=M_Z^2$. This
relation between $v_1^2+v_2^2$ and $M_Z^2$ will no longer be valid
in models with an extended Higgs sector if other fields apart from
the two doublets $H_1$ and $H_2$ contribute with their VEVs to $M_Z^2$
({\em e.g.} triplets). If this happens, then $(g^2+g'^2)v^2/2 < M_Z^2$
and the bound will be more restrictive.

{\bf 3.2 Models with an extended gauge sector.}

In a similar manner, an upper bound on the LHB mass
can be obtained in Supersymmetric models which gauge group (at low
energy, that is $\sim 1\ TeV$) is different from the Standard one
\cite{eq,comelli,comelli2,huitu}. Usually, this type of models require
the introduction of extra representations in the Higgs sector
to give a correct gauge symmetry breaking and of extra fermions
to cancel anomalies. The influence of the extra Higgs
representations on the lightest Higgs bound has been considered in
the previous subsection while the presence of extra exotic fermions
can affect the bound through radiative corrections \cite{moroi}.
In particular
models all these effects have to be combined. In this subsection we
concentrate in the modifications of the tree level bound when the
two doublets $H_1, H_2$
transform non trivially under the extra gauge groups.

Examining the $D$ terms the following contributions to the $\lambda_{ij}$
couplings in (\ref{vstree}) are obtained:
\begin{eqnarray}
\label{deltalg}
&{\displaystyle \delta\lambda_{11}=\frac{1}{2}\sum_a g_a^2(T_1^a)^2\:,\:\:\:
\delta\lambda_{22}=\frac{1}{2}\sum_a g_a^2(T_2^a)^2},&\nonumber\\
&{\displaystyle \delta\lambda_{12}=-\sum_a g_a^2 T_1^aT_2^a},&
\end{eqnarray}
with
\be
\label{tia}
T_i^a\equiv \frac{\langle H_i T^a H_i^*\rangle}{\langle H_i H_i
\rangle},\:\:(i=1,2) .
\ee
This translates in the following modification of the bound on
$m_h^2$ \cite{eq}:
\be
\label{cotag}
\Delta m_h^2 = 2 v^2 \sum_a g_a^2 \left[
T_1^a\cos^2\beta - T_2^a\sin^2\beta\right]^2.
\ee
As a check, for $SU(2)_L$
\be
T_1^L=T_2^L=\frac{1}{2},
\ee
gives $\Delta m_h^2=(g^2v^2/2)\cos^22\beta$. And for $U(1)_Y$
\be
T_1^Y=T_2^Y=\frac{1}{2},
\ee
so that $\Delta m_h^2=(g'^2v^2/2)\cos^22\beta$, in agreement with
the MSSM bound.

After the general discussion in this section it should
be clear that one can apply the bound (\ref{cotah}) also to the SSMM.
Making use of the field rotation (\ref{rot}) directly in the superpotential
(\ref{ssmmw}) we get the term
\be
h^{\nu}_{ij} L_i\cdot H_2 N_j =h^{\nu}_{ij}\frac{x_i}{v'_1} N_j H'_1{}^0
H_2^0 N_j^0 + ... ,
\ee
{\it i.e.} a $\lambda_0$-type coupling
which, according to (\ref{fh}) and (\ref{cotah}) contributes to the bound with
\be
\Delta m_h^2 = v'^2\sum_j \parl\sum_i h^{\nu}_{ij}\frac{x_i}{v'_1}\parr^2
\sin^2 2\beta'.
\ee
This is the same result that was obtained by the direct calculation.
Concerning the $D$ term contribution to the bound, note that the effect
of the field rotation (\ref{rot}) on $V_D$ is
\bear
V_D&=&\frac{1}{8}G^2 \left[\left|H_1^0\right|^2 - \left|H_2^0\right|^2 +
\sum_i\left|\tilde{\nu_i}^0\right|^2 \right]^2\nonumber\\
&=&\frac{1}{8}G^2 \left[\frac{v_1^2}{v'_1{}^2}\left|H_1^0\right|^2
- \left|H_2^0\right|^2 +
\sum_i\frac{x_i^2}{v'^2_1}\left|\tilde{\nu_i}^0
\right|^2 \right]^2+...\nonumber\\
&=&\frac{1}{8}G^2 \left[\left|H'_1{}^0\right|^2 - \left|H_2^0\right|^2
\right]^2+...,
\eear
which is formally equivalent to the MSSM result with the replacement
$H_1{}^0\rightarrow H'_1{}^0$ and so gives the contribution
\be
\Delta m_h^2 = \frac{1}{2}G^2(v'_1{}^2+v_2^2)\cos^22\beta',
\ee
in agreement with (\ref{ssmmb}).
\vspace{1cm}

{\bf 4.} In conclusion, we have improved the tree-level upper bound on
the mass
of the lightest Higgs boson in the Supersymmetric Singlet Majoron Model,
finding a new bound which is controlled by the electroweak scale and
remains light in the limit of heavy exotic VEVs or soft breaking masses
that decouple from the bound.
We have also proved (computing the most important one-loop corrections to
this bound) that this decoupling is not spoiled by radiative corrections.

The similarity of the improved bound calculated in this paper for the
lightest Higgs boson mass in the Supersymmetric Singlet Majoron Model
with previous bounds derived for general Supersymmetric models with an
extended Higgs sector has motivated the re-analysis of the derivation
of these general bounds in order to clarify its
range of applicability. We have shown that those general bounds are in
fact based on very general assumptions, namely that the model contains
a pair of doublets participating in the electroweak breaking and no
fields in $SU(2)_L$ representations higher than triplets take a VEV. With
this simple starting input and using gauge symmetry and supersymmetry to
constrain the effective potential one is able to obtain a bound on the
mass of the lightest Higgs boson of the theory. As a particular example
we have re-derived the bound in the Supersymmetric Singlet Majoron Model
using these general results.

It would be interesting to study whether the decoupling of exotic scales in
one-loop radiative corrections to this tree-level bound is general, that is,
to see if it is automatic for the class of models in which the tree level
bound applies or if further assumptions are required. This subject is
currently under investigation \cite{coes}.
\vspace{1cm}

{\bf Acknowledgements}

It is a pleasure to thank Mariano Quir\'os for enlightening discussions
and a careful reading of the manuscript.

\end{document}